\documentclass[fleqn,usenatbib]{mnras}

\usepackage{newtxtext,newtxmath}
\usepackage{adjustbox}
\usepackage{dcolumn}
\usepackage[T1]{fontenc}
\usepackage{pdflscape}
\DeclareRobustCommand{\VAN}[3]{#2}
\let\VANthebibliography\thebibliography
\def\thebibliography{\DeclareRobustCommand{\VAN}[3]{##3}\VANthebibliography}

\usepackage{graphicx}	
\usepackage{amsmath}	

\title[]{Dynamical substructures of local metal-poor halo}

\author[Dashuang Ye et al.]{
Dashuang Ye,$^{1}$
Cuihua Du,$^{1}$\thanks{E-mail: ducuihua@ucas.ac.cn}
Jianrong Shi$^{2,1}$
and Jun Ma$^{2,1}$
\\
$^{1}$College of Astronomy and Space Sciences, University of Chinese Academy of Sciences, Beijing 100049, P.R. China\\
$^{2}$Key Laboratory of Optical Astronomy, National Astronomical Observatories, Chinese Academy of Sciences, Beijing 100012, P.R.China\\
}

\date{Accepted XXX. Received YYY; in original form ZZZ}

\pubyear{2015}

\begin{document}
\label{firstpage}
\pagerange{\pageref{firstpage}--\pageref{lastpage}}
\maketitle

\begin{abstract}
Based on 4,\,098 very metal-poor (VMP) stars with 6D phase-space and chemical information from \textit{Gaia} DR3 and LAMOST DR9 as tracers, we apply an unsupervised machine learning algorithm, Shared Nearest Neighbor (SNN), to identify stellar groups in the action-energy (\textbf{\textit{J}}-$E$) space. 
We detect seven previously known mergers in local samples, including Helmi Stream, Gaia-Sausage-Enceladus (GSE), Metal-weak Thick Disk (MWTD), Pontus, Wukong, Thamnos, and I'itoi+Sequoia+Arjuna. 
According to energy, we further divide GSE and Wukong into smaller parts to explore the orbital characteristics of individual fragments.
Similarly, the division of Thamnos is based on action.
It can be found that the apocentric distances of GSE parts of high and medium energy levels are located at $29.5\pm3.6\,{\rm kpc}$ and $13.0\pm2.7\,{\rm kpc}$, respectively, which suggests that GSE could account for breaks in the density profile of the Galactic halo at both $\approx30$\,kpc and $15\text{-}18$\,kpc. 
The VMP stars of MWTD move along prograde orbits with larger eccentricities than those of its more metal-rich stars, which indicates that the VMP part of MWTD may be formed by accreting with dwarf galaxies.
Finally, we summarize all substructures discovered in our local VMP samples. 
Our results provide a reference for the formation and evolution of the inner halo of the Milky Way (MW).
\end{abstract}




\section{Introduction}
The proto-MW has undergone frequent mergers with small progenitor galaxies.
According to the $\rm\Lambda CDM$ cosmological model, the MW forms through hierarchical processes \citep{white1991}, which predicted that the MW involves a series of accretion events.
Many studies have shown that the Galactic stellar halo is primarily formed by the merging of numerous progenitor galaxies \citep{helmi1999,majewski2003,newberg2009,helmi2018merger,belokurov2018co,myeong2018sausage,koppelman2019multipleretrograde,myeong2019,yuan2020lowmass,naidu2020,naidu2021,malhan2022}.
The tidal disruption of the merging galaxy occurs slowly enough to form a vast stellar stream in the Galactic halo \citep[e.g., this is the case for the Sagittarius merger;][]{Ibata2020,vasiliev2020sgr}.
However, for radially accreting galaxies, the stars will quickly get phase-mixed and no clear signature of the stream will be visible \citep[e.g., Gaia-Sausage/Enceladus][]{belokurov2018co,helmi2018merger}.
Fortunately, they retain some of the orbital and chemical characteristics of their progenitor galaxies.
Therefore, in order to explore the tidal debris from their progenitor galaxies, a common approach is clustering in the integrals of motion space (e.g., actions).

Massive amounts of all-sky high-quality photometry and astrometry have been provided by the \textit{Gaia} Data Releases \citep{Gaia2016,Gaiadr3}.
For more complete information, its precise proper motion and parallax can be combined with chemical information and radial velocities from other spectroscopic surveys, such as LAMOST \citep{cui2012large}, APOGEE \citep{majewski2017,APOGEE2022ApJS}, RAVE \citep{RAVE12020AJ,RAVE22020AJ}, SEGUE \citep{yanny2009AJ,Alam2015ApJS}, GALAH \citep{desilva2015,martell2017MNRAS,galah2021MNRAS}.
Thanks to these large stellar spectroscopic surveys, the combination of motion and chemical abundances has become available for millions of stars in the solar neighbourhood, which allows us to detect substructures in the local halo.

Many studies have focused on identifying and describing the MW's stellar populations, and try to determine whether they originated from the MW (i.e., in-situ halo) or nearby dwarf galaxies.
Detection of substructure in phase space has worked successfully for the identification of accretion events.
\citet{helmi1999} found that the Helmi Streams originate from a dwarf galaxy of $\sim10^8~\text{M}_{\odot}$ that was accreted $5\text{-}8~{\rm Gyr}$ ago \citep{helmi2006,koppelman2019characterization}.
Other prominent discoveries are the core of Sagittarius (Sgr) dwarf galaxy \citep{ibata1994,ibata1995,yanny2000,vasiliev2020sgr} and it is still forming tidal stream \citep{ibata2001,majewski2003,belokurov2014,Hernitschek2017,ramos2020}, which have been used as a paradigm for how dwarf galaxies merger with the MW.
The majority of the inner halo comes from the merger of a massive dwarf galaxy with an initial stellar mass of $5~\times~10^8-5~\times~10^9~{\rm M_{\odot}}$ known as Gaia-Sausage-Enceladus \citep[][]{belokurov2018co,helmi2018merger,Haywood2018GS,mackereth2019,vincenzo2019}, it was accreted by the MW at $z\approx2$ \citep[$\sim10~{\rm Gyr}$ ago,][]{dimatteo2019,gallart2019}.
Furthermore, in recent years, several phase-space substructures believed to be remnants of merged galaxies have been identified in the Galactic stellar halo, including Sequoia \citep{myeong2018discovery,myeong2019}, Thamnos \citep{koppelman2019multipleretrograde}, Wukong/LMS-1 \citep{yuan2020lowmass,naidu2020}, I'itoi, Arjuna, Aleph \citep{naidu2020}, Heracles \citep{Horta2021}, Icarus \citep{ReFiorentin2021}, Cetus \citep{newberg2009,Thomas2022A&A}, and Pontus \citep{malhan2022}.
In addition, \citet{wang2022} used the friends-of-friends algorithm to identify substructures in five-dimensional space, i.e., eccentricity $e$, semimajor axis $a$, the direction of the orbital pole $(l_{\rm orbit},b_{\rm orbit})$ and the angle between apocenter and the projection of $x$-axis on the orbital plane $l_{apo}$, and found three remaining unknown substructures and one of them has large angular momentum and a mean metallicity -2.13 dex.

The MW's potential presumably evolved adiabatically.
Considering that the actions \textbf{\textit{J}} are conserved for a long time if the potential evolved adiabatically, those stars merged from a progenitor galaxy would be clumped in actions \textbf{\textit{J}} space.
Many recent studies \citep[e.g.,][]{li2019substructures,koppelman2019multipleretrograde,yuandynamicalrelics2020,litwosubstructures,naidu2020,malhan2022} refer to energy $E$ as an extra quantity to distinguish the mergers even if it is not adiabatic, and their results are highly reminiscent of the substructures from numerical simulations of the disruption of satellite galaxies by the Galactic potential \citep{helmi2000mapping,Gomez2013MNRAS,Orkney2023MNRAS}.

In this paper, we use the two-stage SNN to identify dynamical relics in the VMP local halo. In Section \ref{Data}, we describe the selection criteria and dataset. We introduce the clustering method in Section \ref{Method}, and then analyze all the substructures in Section \ref{Results}. Finally, conclusions are given in Section \ref{Conclusion}.

\section{Data}\label{Data}
In this paper, we use proper motions from \textit{Gaia} DR3 \citep{Gaia2016, Gaiadr3} as well as chemical information (i.e., [Fe/H], $\rm [\alpha/Fe]$) and line-of-sight velocities from LAMOST DR9 \citep{cui2012large,zhao2012lamost,liu2020lamost}.
We apply selection cuts for \textit{Gaia} DR3:
\begin{itemize}
  \item {\tt\string parallax}$\,\geqslant\,0.2\,{\rm mas}$
  \item $\cfrac{\tt\string parallax\underline{~}error}{\lvert\tt\string parallax\rvert}\,\leqslant\,0.2$
  \item ${\tt\string ruwe}\,\textless\,1.2$
  \item $6\,{\rm mag}\,\textless\,{\tt\string phot\underline{~}g\underline{~}mean\underline{~}mag}\,\textless\,21\,{\rm mag}$
  \item $1.1\,{\rm \mu m^{-1}}\,\textless\,{\tt\string nu\underline{~}eff\underline{~}used\underline{~}in\underline{~}astrometry}\,\textless\,1.9\,{\rm\mu m^{-1}}\\({\tt\string astrometric\underline{~}params\underline{~}solved=31})$
  \item $1.24\,{\rm\mu m^{-1}}\,\textless\,{\tt\string pseudocolour}\,\textless\,1.72\,{\rm\mu m^{-1}}\\({\tt\string astrometric\underline{~}params\underline{~}solved=95})$
\end{itemize}
where the last three conditions are for reliable parallaxes zero-point, and then cross-match with the LAMOST DR9 catalogue ($\textless\,1^{\prime\prime}$) to obtain metallicities and line-of-sight velocities.
By comparing line-of-sight velocities between \textit{Gaia} DR3 and LAMOST DR9, we notice a systematic bias of 5.32 ${\rm km s^{-1}}$ with a dispersion ${\rm\sigma_{rv}}=7.26\,{\rm km\,s^{-1}}$, which we add to the LAMOST velocities, and remove stars with $\lvert v_{\rm los,\textit{Gaia}}-v_{\rm los, LAMOST}-5.32\rvert\,\textgreater\,3{\rm\sigma_{rv}}$ that may be mismatched or have unreliable line-of-sight velocities.
Finally, we utilize the following selection cuts:
\begin{itemize}
  \item ${\rm radial\ velocity\ uncertainties\ (LAMOST)}\,\textless\,10\,{\rm km\,s^{-1}}$
  \item ${\rm [Fe/H]\ error\ (LAMOST)}\,\textless\,0.2\,{\rm dex}$
  \item ${\rm S/N\ of\ g\ filter\ (LAMOST)}\,\textgreater\,20$
  \item ${\rm [Fe/H]\ (LAMOST)}\,\textless\,-1.8\,{\rm dex}$
  \item $\lvert \textbf{v}-\textbf{v}_{\rm LSR} \rvert \textgreater~210{\rm~km~s^{-1}}$
\end{itemize}
where $v_{\rm los,\textit{Gaia}}$ and $v_{\rm los,LAMOST}$ are the line-of-sight velocities of \textit{Gaia} and LAMOST.
Finally, there are 4098 VMP stars with full 6-D phase-space information.
We apply the method in \citet{bailer2021estimating} to estimate geometric distances, in which we sample the posterior probability with Goodman \& Weare's affine-invariant Markov Chain Monte Carlo \citep[MCMC,][]{goodman2010ensemble}, using the \textbf{\tt\string emcee} python package \citep{foreman2013emcee}.
When calculating the distance, we take into account the correlation coefficient between parallax and proper motion \citep[e.g.,][]{ducuihua2019,yanyepeng2020,litwosubstructures,zhuhaifan2021,liaojiwei2023}.

We adopt a right-handed Galactocentric frame of reference similar to the one defined in \citet{li2019substructures}: here $x$, $y$ and $z$ indicate the Cartesian coordinates; $R$ is the cylindrical radius, $r$ is the spherical radius, and $\phi$ and $\theta$ represent the azimuthal and zenithal angle, respectively.
In this coordinate system, the Sun is located at (8.2, 0, 0.025)\,kpc \citep{juric2008milky,bland2016galaxy}. 
We take $V_{\rm LSR}=232.8\,{\rm km\,s^{-1}}$ \citep{mcmillan2016mass} for the local standard of rest (LSR) and $(U_{\odot},\,V_{\odot},\,W_{\odot})=(10,\,11,\,7)\,{\rm km\ s^{-1}}$ \citep{tian2015stellar,bland2016galaxy,mcmillan2016mass} for the Sun's proper motion with respect to the LSR.

In this study, we use the \citet{mcmillan2016mass} model because the predicted velocity curve of this model is more consistent with the measurements of the Milky Way \citep[e.g.,][]{bovy2020,Nitschai2021}.
To compute actions (\textbf{\textit{J}}), energy ($E$), and other orbital parameters, we make use of the {\tt\string galpy} module \citep{bovy2015}.
We analyze actions ($J_R,\ J_{\phi},\ J_z$) in cylindrical coordinates system, where $J_{\phi}$ is equal to the $z$ component angular momentum ($L_z$) under \citet{mcmillan2016mass} model, and a negative $J_{\phi}$ represents prograde motion.
$J_R$ and $J_z$ describe the extent of oscillations in cylindrical radius and $z$ directions, respectively.
In addition, we obtain other orbital parameters, including eccentricity, apocentric distance, pericentric distance and the maximum vertical height ($e,\ r_{\rm apo},\ r_{\rm peri},\ z_{\rm max}$).
For each star, we construct a set of 100 initial conditions using a Monte Carlo technique considering the observational uncertainties in heliocentric distances, proper motions, and line-of-sight velocities.
The final dynamical parameters are taken as the means of the derived distributions, and associated uncertainties are the corresponding standard deviations.
Note that we calculate the $4\times4$ covariance matrices about energy and actions by {\tt\string biweight\underline{~}midcovariance} in the Python module {\tt\string astropy} \citep{astropy2013A&A} because it can remove spurious outliers.

\begin{figure}
	\includegraphics[width=1\columnwidth]{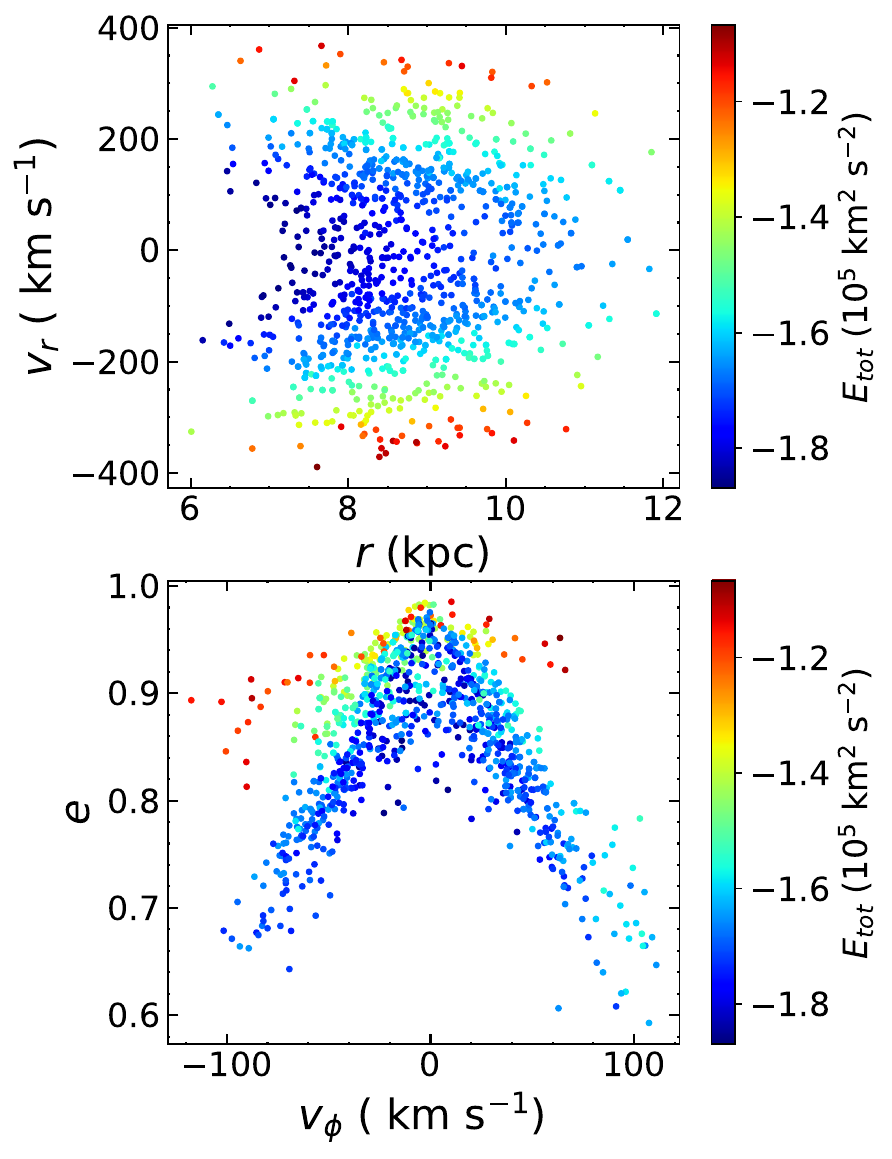}
    \caption{The member stars in a massive substructure, namely GSE, are in the $r\text{--}v_r$ space (top panel) and $v_{\phi}\text{--}e$ space (bottom panel) and colour-coded according to energy.}
    \label{figGSE_fold}
\end{figure}

\begin{figure*}
	\includegraphics[width=0.9\textwidth]{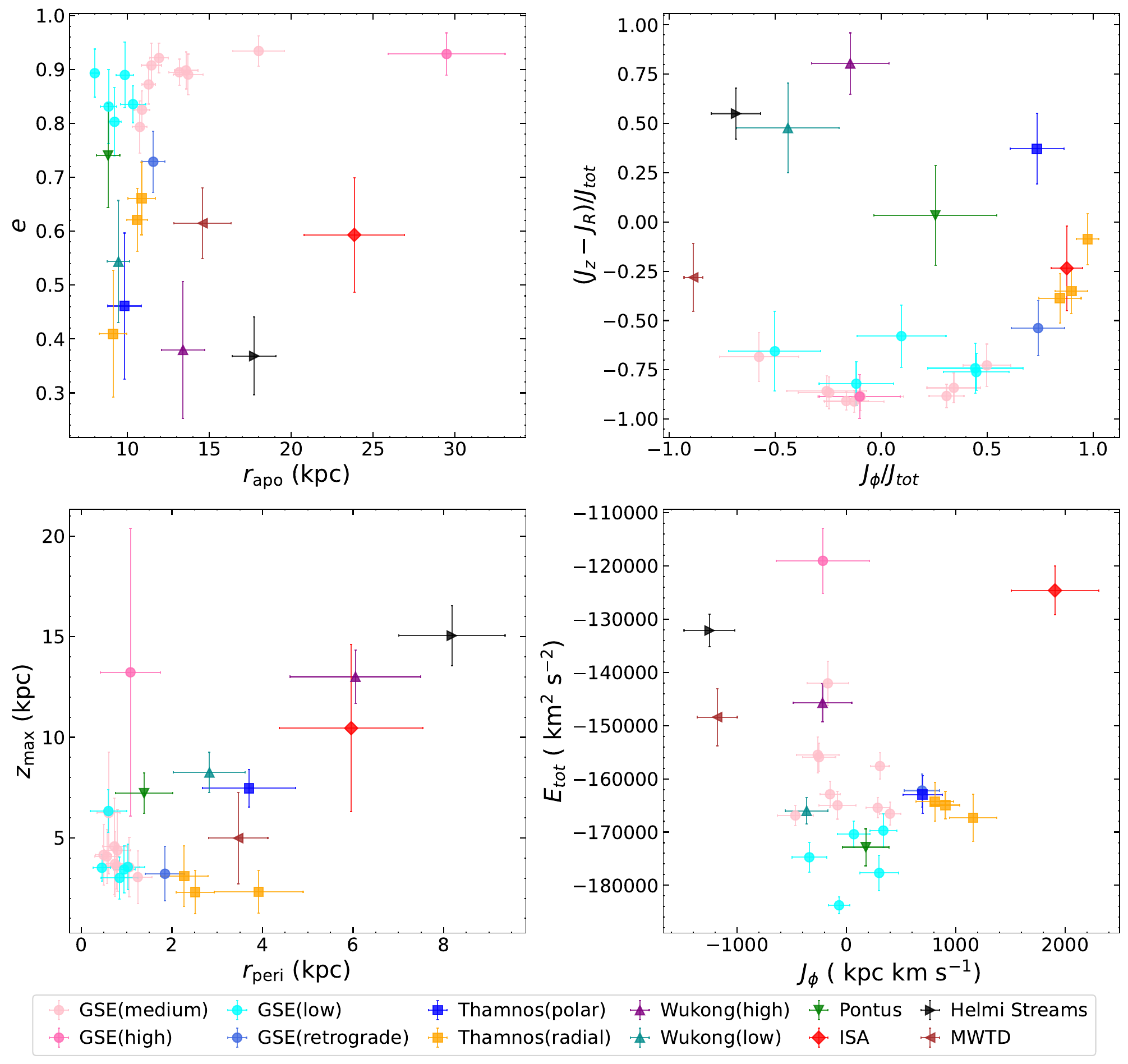}
    \caption{26 groups assigned to seven substructures plotted by distinct shapes and colour where error bars in each object are the standard deviations of the \textit{x}- and \textit{y}-axis quantities. For the subdivided substructures (e.g., GSE, Thamnos, Wukong), each part is represented in a different colour but with the same shape. All groups are shown in $r_{\rm apo}\text{--}e$ space (top-left), projected action space (top-right), $r_{\rm peri}\text{--}z_{\rm max}$ space (bottom-left), and $E\text{--}J_{\phi}$ space (bottom-right).}
    \label{figgroups_errorbar}
\end{figure*}

\begin{figure*}
	\includegraphics[width=1\textwidth]{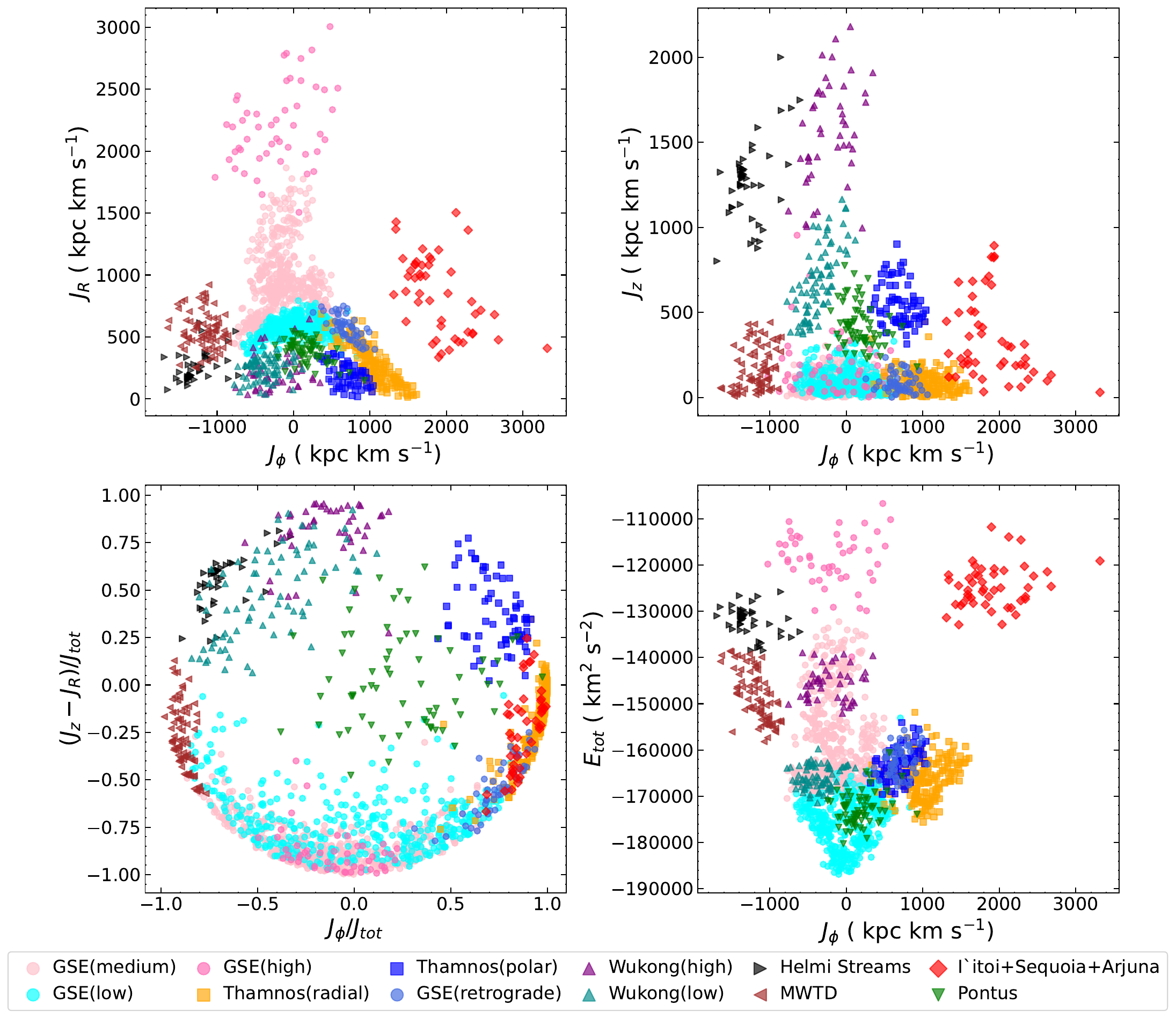}
    \caption{Seven substructures VMP stars described in the (\textit{\textbf{J}},\,\textit{E}) space, using the same colour and shape scheme as Figure \ref{figgroups_errorbar}. Note that the retrograde GSE part (royal blue circle) coincides with the polar Thamnos (blue square) in the $E\text{--}J_{\phi}$ space, but does not happen in the projected action space (bottom-left).}
    \label{figallgroups}
\end{figure*}

\begin{figure*}
	\includegraphics[width=1\textwidth]{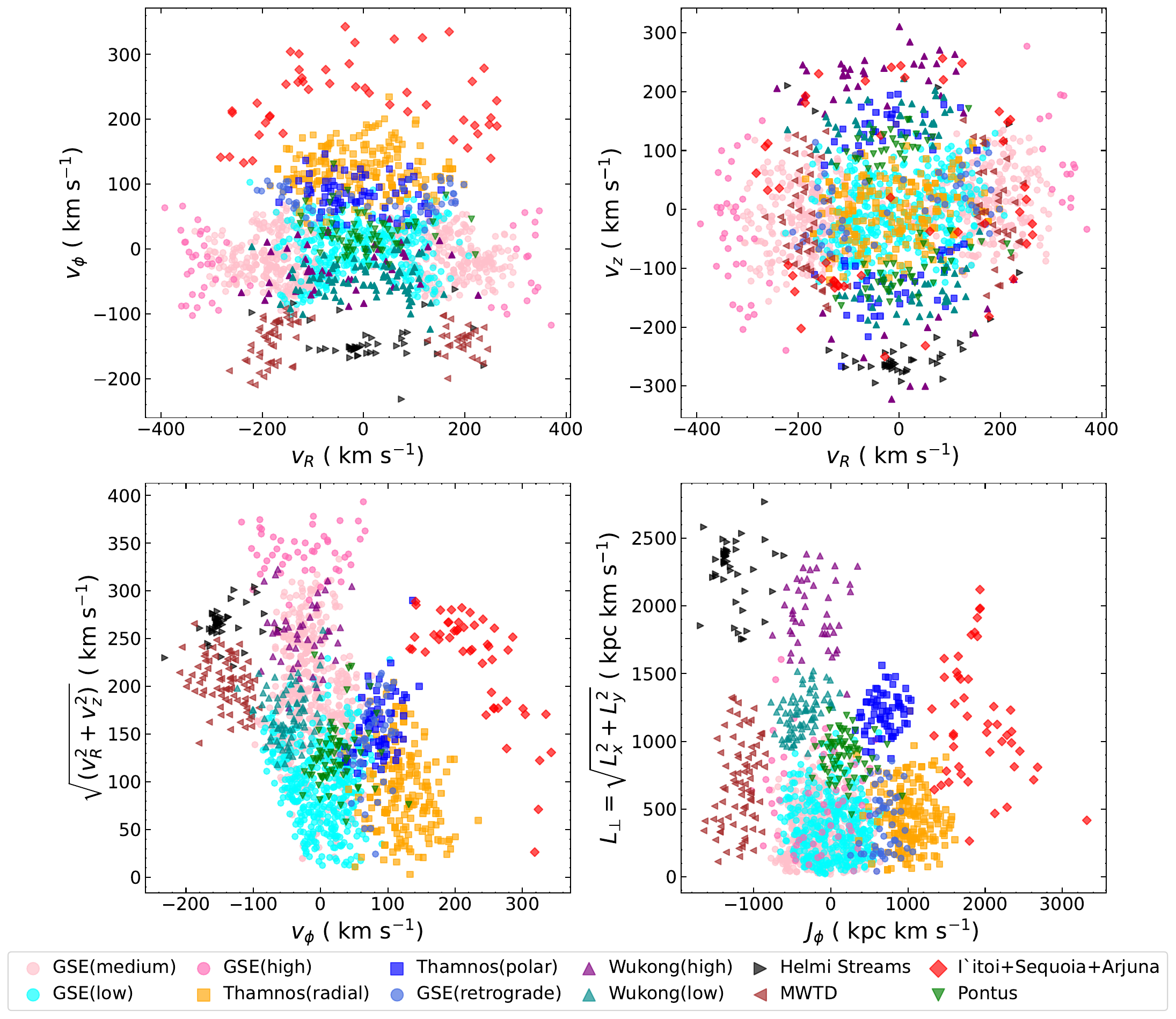}
    \caption{3D kinematics in cylindrical coordinates for all substructures, using the same colour and shape scheme as Figure \ref{figgroups_errorbar}.}
    \label{figallgroups_v}
\end{figure*}

\begin{figure*}
	\includegraphics[width=0.96\textwidth]{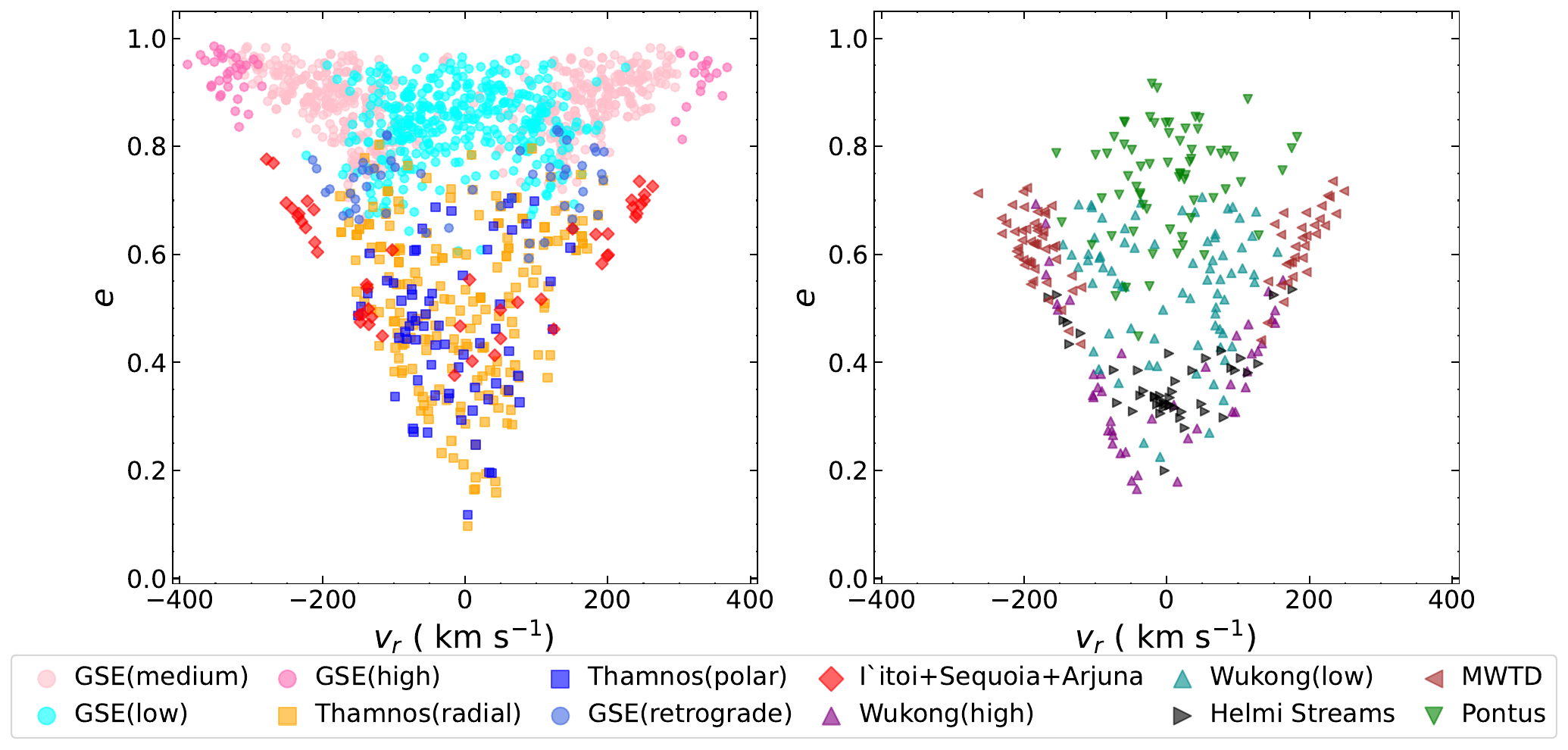}
    \caption{\textit{Left:} GSE, Thamnos and ISA in the $v_r\text{--}e$ space. \textit{Right:} Wukong, Pontus, Helmi Streams and MWTD in the $v_r\text{--}e$ space. Note that we present all substructures in two separate panels solely to avoid visual clutter.}
    \label{figallgroups_evr}
\end{figure*}

\begin{figure*}
	\includegraphics[width=1\textwidth]{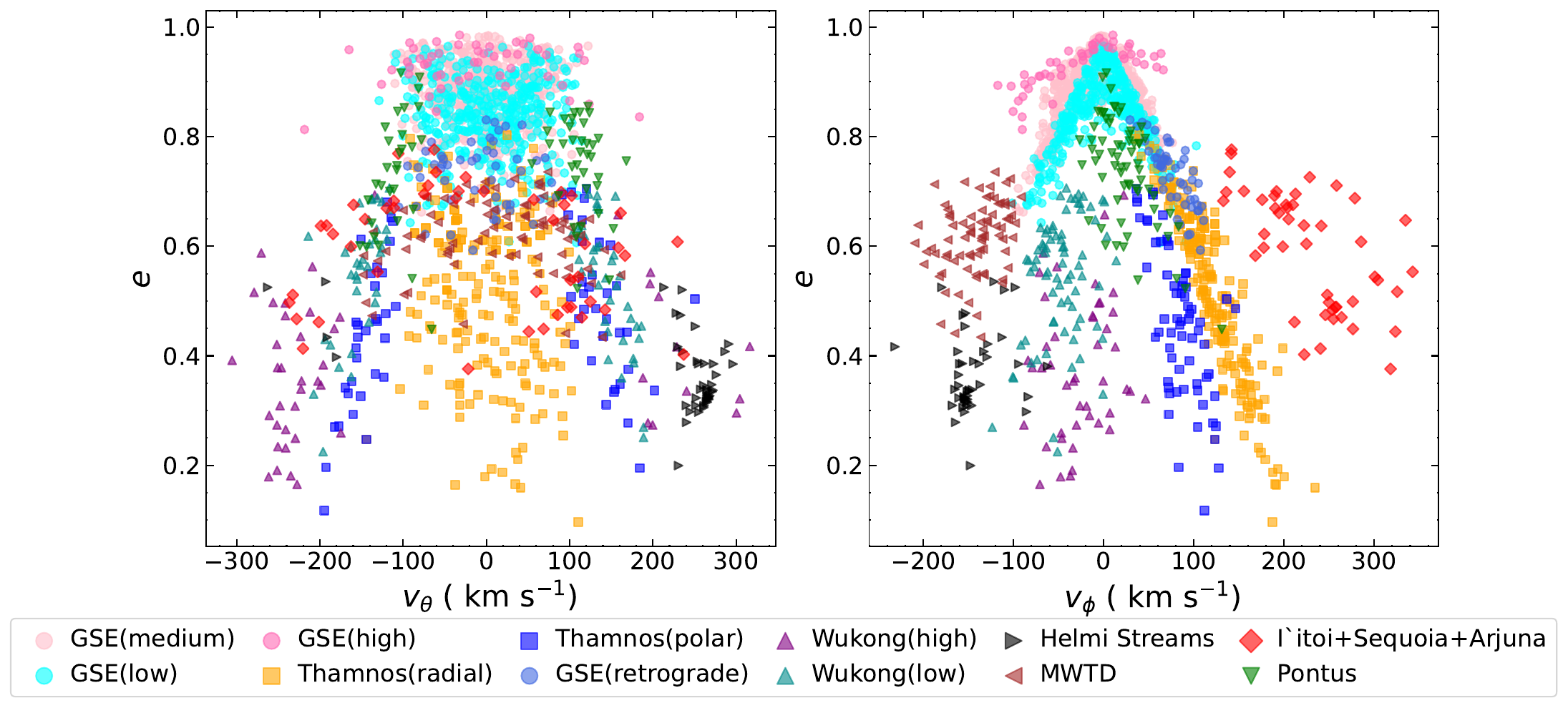}
    \caption{All substructures in the $v_{\theta}\text{--}e$ space (left panel) and $v_{\phi}\text{--}e$ space (right panel).}
    \label{figallgroups_evt}
\end{figure*}

\section{Method}\label{Method}
In this work, we search for structure among the VMP stars from LAMOST DR9 in the energy-action space.
Energy is conserved as long as the potential of the Milky Way is static, and the actions are insensitive to the slow, adiabatic time-dependence of the potential.
Halo stars belonging to the same structure, even when they are scattered across the sky, retain similar coordinates in the energy-action space, so they could be revealed through clustering algorithms.

In this section, we summarize the method of two-stage SNN clustering; further details can be found in \citet{chen2020discovery}.
The SNN \citep{ertoz2003finding} can be seen as a modified version of DBSCAN, short for density-based spatial clustering of applications with noise \citep{ester1996density}. 
The SNN adopts the Jaccard distance metric of the nearest neighbours in \textbf{\textit{J}} space to make the density threshold more flexible.
The Jaccard distance is defined as:
\begin{equation}
\begin{aligned}
d_{\rm Jaccard}(A,\,B) = 1 - \frac{\lvert S_{A} \cap S_{B} \rvert}{\lvert S_{A} \cup S_{B}\rvert}.
\end{aligned}
\end{equation}
where $S_{A}$ and $S_{B}$ represent the sets of the nearest neighbors in \textbf{\textit{J}} space for two stars, $A$ and $B$. $\lvert S_{A}\cap S_{B}\rvert$ and $\lvert S_{A}\cup S_{B}\rvert$ represent the sizes of the intersection and union of the nearest neighbours of two stars, respectively.
Thus, for those sets that are completely overlapping, $d_{\rm Jaccard}$ equals 0, whereas for those sets that are entirely disjoint, $d_{\rm Jaccard}$ equals 1.

The first stage consists of three steps: retrieve the same number ($\rm N_{\textbf{\textit{J}}}$) of nearest neighbours for each star in actions space, and then remove those with dissimilar energies ($\textgreater\,d_{E}$) in order to keep neighbours that share both similar actions and energies;
compute the Jaccard distance matrix; cluster by DBSCAN implemented in Python module {\tt\string scikit-learn}\footnote{\url{https://scikit-learn.org/stable/user_guide.html}} \citep{sklearn2014}.
Two free parameters are involved in first-stage clustering: one is the number (${\rm N}_{\textbf{\textit{J}}}$) of the nearest neighbours for each star in the action space, and the other is the maximum energy difference (${\rm d}_{E}$) between a star and its nearest neighbours in actions space.
For a large $\rm N_{\textbf{\textit{J}}}$, the elements of the Jaccard matrix tend to be larger, which, in turn, increases the sizes of groups composed of stars with similar actions. 
However, such a small set of the nearest neighbour leads to a weak correlation among most of the stars, consequently resulting in very few or even no clusters.
In addition, $d_{E}$ is a secondary parameter, as it imposes energy constraints on the nearest neighbours composed of $\rm N_{\textbf{\textit{J}}}$ members.
Excessive $d_E$ values loosen these constraints, while extremely small values significantly decrease the size of the nearest neighbour.
Therefore, adjusting these parameters can indeed be a formidable challenge.

We apply a Monte Carlo approach to accomplish parameter tuning and calculate the frequentist stability of groups (the number of times a group appears) and the probability of members belonging to an assigned group (the percentage of times a star appears in the duplicates of an assigned group).
First of all, the maximum possible value of $d_E$ should not be set too high initially, because some adjacent substructures in the $E_z\text{-}L_z$ space exhibit small energy differences, such as Wukong and GSE. Additionally, in cases where $\rm N_{\textbf{\textit{J}}}$ is too large, it usually leads to stars that are not strongly correlated in action space being assigned to a common group.
Similar to \citet{chen2020discovery}, we set the ranges of ${\rm N}_{\textbf{\textit{J}}}$ and ${\rm d}_{E}$ to $20-220$ (each star has a chance to connect to at most about 5\% of the sample closest to it in the action space) and $0-6600\,{\rm km^{2}\,s^{-2}}$ (twice the average error of energy), and then redraw 10,000 parameter values from uniform distributions, respectively.
Taking into measurement errors, we resample the energies and actions of our sample stars according to their $4\times4$ covariance matrices about energy and actions.
We ran the SNN algorithm 10,000 times with randomly generated samples and parameter values, and we obtained 41,229 stellar groups.
Of course, not all of these $\sim40,000$ groups are reliable, so we need to cluster these groups, regarding each group as an individual, in order to identify high-frequency or stable groups.

In the second stage of clustering, we calculate the Jaccard matrix using these stellar groups instead of the nearest neighbour of each star in \textbf{\textit{J}} space, and then apply DBSCAN to cluster again.
In order to remove stars assigned to, with low probabilities, groups as well as to exclude groups that appear by chance due to specific $\rm N_{\textbf{\textit{J}}}$ and $d_E$ values, we must set probability and stability thresholds. 
To avoid confusion from fortuitous overlaps, we select stars with stability (the number of times a group appears) above 30 and a probability (the percentage of times a star appears in the duplicates of an assigned group) $\textgreater\,30\%$ so that only a few stars are assigned to multiple groups.
Even for the group with a stability of 31, the probability of its members ($\sim73\%$) is much larger than the probability threshold ($30\%$), indicating that the number of its members is not large enough to be significantly recognized multiple times but it is highly compact.
For a group composed of members with a probability of $\sim30\%$, its stability is outstanding.
Finally, we summarize 26 groups with 1,\,515 stars, as listed in Table \ref{tab:VMP groups}, simultaneously removing 2,\,583 stars that probabilistically ( $\rm probability\,\textless\,30\%$ ) belong to poor-stability ( $\rm stability\,\textless\,30$ ) multiple groups and represent smooth, diffuse VMP background halo without clumping in our 4D phase space.
Smooth stellar halo is defined as background halo obtained by removing obvious substructures. 
It is utilized for measuring the anisotropy profile \citep{bird2021}, fraction of GSE \citep{wuwenbo2022}, density shape \citep{Wu2022AJ}, and Milky Way mass \citep{bird2022MNRAS}.
Smooth VMP background halo stars also involve member stars from ancient mergers, but further determination of their origins requires detailed chemical information or model fitting.
\begin{table}
 \caption{Summary of VMP groups.}
 \label{tab:VMP groups}
 \begin{tabular*}{\columnwidth}{lllll}
  \hline
  Group$^a$ & \textit{n}$^b$ & Stability & Probability$^c$ & Substructure$^d$ \\
  \hline
  G-1 & 104 & 5446 & $0.414\pm0.075$ & GSE (pink) \\
  G-2 & 45 & 3492 & $0.800\pm0.239$ & Helmi Streams (black) \\
  G-3 & 80 & 1243 & $0.431\pm0.090$ & MWTD (brown)\\
  G-4 & 32 & 1034 & $0.370\pm0.056$ & GSE (pink) \\
  G-5 & 66 & 490 & $0.535\pm0.206$ & Thamnos (blue) \\
  G-6 & 45 & 240 & $0.701\pm0.213$ & Wukong/LMS-1 (purple) \\
  G-7 & 75 & 208 & $0.476\pm0.131$ & Wukong/LMS-1 (dark cyan) \\
  G-8 & 54 & 185 & $0.513\pm0.163$ & GSE (pink) \\
  G-9 & 46 & 169 & $0.413\pm0.083$ & GSE (pink) \\
  G-10 & 47 & 169 & $0.577\pm0.198$ & ISA (red)\\
  G-11 & 62 & 133 & $0.547\pm0.163$ & GSE (cyan) \\
  G-12 & 59 & 121 & $0.491\pm0.156$ & GSE (pink) \\
  G-13 & 89 & 60 & $0.668\pm0.190$ & GSE (pink) \\
  G-14 & 50 & 59 & $0.585\pm0.201$ & Thamnos (orange) \\
  G-15 & 85 & 52 & $0.379\pm0.060$ & GSE (cyan) \\
  G-16 & 52 & 52 & $0.596\pm0.219$ & GSE (royal blue) \\
  G-17 & 49 & 51 & $0.609\pm0.227$ &  GSE (hot pink) \\
  G-18 & 108 & 51 & $0.409\pm0.096$ &  Thamnos (orange) \\
  G-19 & 47 & 48 & $0.645\pm0.234$ &  GSE (cyan) \\
  G-20 & 59 & 47 & $0.597\pm0.210$ &  GSE (pink) \\
  G-21 & 81 & 45 & $0.538\pm0.181$ &  GSE (pink) \\
  G-22 & 61 & 40 & $0.695\pm0.182$ &  GSE (cyan) \\
  G-23 & 67 & 39 & $0.597\pm0.235$ &  Pontus (green)\\
  G-24 & 71 & 39 & $0.516\pm0.170$ &  GSE (pink) \\
  G-25 & 119 & 37 & $0.471\pm0.107$ &  GSE (cyan) \\
  G-26 & 50 & 31 & $0.726\pm0.214$ &  Thamnos (orange) \\
  \hline
  \multicolumn{5}{l}{$^a$ Group represents the name of cluster derived from the two-stage SNN.}\\
  \multicolumn{5}{l}{$^b$ \textit{n} represents the size of cluster.}\\
  \multicolumn{5}{l}{$^c$ it is the mean and standard deviation of probability for stars associated}\\
  \multicolumn{5}{l}{with each group.}\\
  \multicolumn{5}{l}{$^d$ The contents in parentheses indicate colours coded in Figures \ref{figgroups_errorbar}-\ref{figallgroups_evt}.}\\
 \end{tabular*}
\end{table}

Considering that for some mergers, their member stars could be multiply stacked in the 4D phase space, such as energy wrinkles and phase-space folds from GSE \citep{Belokurov2023MNRAS,Wu2023ApJ}, we cannot simply think that each group corresponds to a distinct and individual merger.
Based on orbital properties, we divided the 26 groups into seven substructures, including Gaia-Sausage-Enceladus, Thamnos, Metal-weak Thick Disc, Helmi Streams, Wukong/LMS-1, I'itoi+Sequoia+Arjuna, Pontus, as follows:
\begin{itemize}
 \item Gaia-Sausage-Enceladus: $(e\,\textgreater\,0.7)\ \wedge\ ((J_z-J_R)/J_{\rm tot}\,\textless\,-0.5)$,
 \item  Thamnos: $(0.6\,\textless\,J_{\phi}/[{\rm 10^3\,kpc\,km\,s^{-1}}]\,\textless\,1.2)\ \wedge\ (-1.7\,\textless\,E/[{\rm 10^5\,km^2\,s^{-2}}]\,\textless\,-1.6)$\ $\wedge$\ (excluding all previously defined structures),
 \item  Metal-weak Thick Disc: $(\lvert(J_z-J_R)/J_{\rm tot}\rvert\,\textless\,0.3)\ \wedge\ (J_{\phi}/J_{\rm tot}\,\textless\,-0.8)$\ $\wedge$\ (excluding all previously defined structures),
 \item  Helmi Streams: $(-1.7\,\textless\,L_z/[{\rm 10^3\,kpc\,km\,s^{-1}}]\,\textless\,-0.75)\ \wedge\ (1.6\,\textless\,L_{\bot}/[{\rm 10^3\,kpc\,km\,s^{-1}}]\,\textless\,3.2)$\ $\wedge$\ (excluding all previously defined structures),
 \item  Wukong/LMS-1: $(J_{\phi}/J_{\rm tot}\,\textless\,0)\ \wedge\ ((J_z-J_R)/J_{\rm tot}\,\textgreater\,0)$\ $\wedge$\ (excluding all previously defined structures),
 \item  I'itoi+Sequoia+Arjuna: $(E/[{\rm 10^5\,km^2\,s^{-2}}]\textgreater-1.3)\wedge(J_{\phi}/[{\rm 10^3\,kpc\,km\,s^{-1}}]\,\textgreater\,1.4)$\ $\wedge$\ (excluding all previously defined structures),
 \item  Pontus: $(0.5\,\textless\,e\,\textless\,0.8)\ \wedge\ (0.245\,\textless\,J_R/[{\rm 10^3\,kpc\,km\,s^{-1}}]\,\textless\,0.725)\ \wedge\ (-0.005\,\textless\,J_{\phi}/[{\rm 10^3\,kpc\,km\,s^{-1}}]\,\textless\,0.470)\ \wedge \ (0.115\,\textless\,J_z/[{\rm 10^3\,kpc\,km\,s^{-1}}]\,\textless\,0.545)\ \wedge \ (E/[{\rm 10^5\,km^2\,s^{-2}}]\,\textless\,-1.6)$\ $\wedge$\ (excluding all previously defined structures).
\end{itemize}
Note that we select substructures by constraining the mean orbital parameters in each group, so a few stars belonging to a substructure cannot satisfy its selection criteria.
In this work, we did not identify previously discovered five substructures, including Sagittarius (Sgr), Heracles, Aleph, Icarus, and Cetus due to: (\romannumeral1) the heliocentric distances of Sgr stars generally are beyond 10\,kpc \citep{Hayes2020ApJ}, while the farthest distance in our sample is just close to 5\,kpc; (\romannumeral2) Heracles is located in the heart of the Galaxy \citep[$r\,\textless\,4\,$kpc,][]{Horta2021}, but all stars in our sample are beyond $r\sim5\,$kpc; (\romannumeral3) Aleph is a metal-rich ($\rm [Fe/H]\,\textgreater\,-0.8$) substructure \citep{naidu2020}; (\romannumeral4) Icarus is composed of stars that are metal-poor ($\rm[Fe/H]\sim-1.45$) with circular orbits \citep{ReFiorentin2021}; (\romannumeral5) Cetus is a diffuse stream orbiting at large heliocentric distances \citep[$d_{\odot}\,\gtrsim\,30\,$kpc,][]{newberg2009,Thomas2022A&A}.

\section{Results}\label{Results}
In this section, for each substructure discussed below, we analyze its dynamical properties $(\textbf{\textit{J}},\,E)$ and orbital parameters. 
In particular, we show all the mergers in ``projected action space'' represented by a diagram of $J_{\phi}/J_{\rm tot}$ versus $(J_z-J_R)/J_{\rm tot}$, where $J_{\rm tot}$ = $\sqrt{J_{R}^{2}+J_{z}^{2}+J_{\phi}^{2}}$.
The reason for using the projected action space is that this plot is effective in separating objects that lie along circular, radial, and in-plane orbits, and it is considered to be superior to other kinematic spaces \citep[e.g.,][]{james2022}.
\subsection{Gaia Sausage/Encelaus}

It has been suggested that the structure known as the Gaia-Sausage-Enceladus (GSE) is the remnant of the Galaxy's last major accretion event \citep{belokurov2018co,helmi2018merger,myeong2018sausage, Haywood2018GS}.
\citet{wuwenbo2022} recently found that the GSE contributed about $41\%-71\%$ of the inner ($r\,\textless30\,{\rm kpc}$) stellar halo by fitting their K giant sample with the Gaussian mixture model.
The GSE members can be distinguished in velocity space, as they form an elongated distribution in $v_R$ around a close-to-zero azimuthal velocity \citep{kpppelman2018}.
In addition, due to their high eccentricities, the GSE stars can be selected using their orbital properties \citep[e.g.,][]{myeong2018sausage,naidu2020, Bonaca2020}.
\citet{naidu2020} defined the GSE as the highly radial population and selected the GSE stars by requiring $e\,\textgreater\,0.7$.
We preliminarily selected 17 groups by $e\,\textgreater\,0.7$, and then removed a slightly retrograde group belonging to Pontus \citep{malhan2022} by $(J_z-J_R)/J_{\rm tot}\textless-0.5$, because the orbits of Pontus stars are more polar than those of GSE stars.

Recently, \citet{Belokurov2023MNRAS} found a series of long and thin chevrons-like overdensities associated with energy in the $(r,\ v_r)$ space, which they interpret as folds of GSE tidal debris as it stretches and winds up due to phase-mixing in the MW gravitational potential.
In addition, they found that the $(E,\ L_z)$ distribution appears more prograde at high energies but is roughly symmetric with respect to $L_z$ at low energies.
Further numerical simulation shows that the least bound population of GSE-like satellite have on average slightly positive $L_z$, inheriting it from the satellite's orbital angular momentum, and more tightly bound debris are mostly located closer to $L_z=0$.
In the top panel of Figure \ref{figGSE_fold}, based on the GSE VMP stars, the chevrons can also be seen, indicating that they are strongly related to energy.
The angle between two wings of chevron decreases with energy, which is consistent with that in \citet{Belokurov2023MNRAS}.
In the bottom panel of Figure \ref{figGSE_fold}, we additionally find that the $(v_{\phi},\ e)$ distribution exhibits the short and wide chevrons, and their angles increase with energy.
The GSE stars are symmetrically distributed relative to $v_{\phi}\sim0$ between $E\sim-1.6$ and $\sim-1.8\,[\times10^5\ {\rm km^2\,s^{-2}}]$, while the GSE stars are inclined to prograde ($v_{\phi}\,\textless\,0$) between $E\sim-1.1$ and $\sim-1.6\,[\times10^5\ {\rm km^2\,s^{-2}}]$.
In the bottom-right of Figure \ref{figgroups_errorbar}, for GSE (pink, hot pink, cyan, and royal blue circles), the $(E,\ L_z)$ distribution appears stratified, corresponding to the energy wrinkles found by \citet{Belokurov2023MNRAS}. In their work, energy wrinkles are only distributed at prograde and retrograde edges, while a continuous, smooth and asymmetric peak appears at low $\lvert L_z\rvert$.
Here, energy wrinkles are tightly presented at low $\lvert L_z\rvert$, probably because what we obtained are only high-stability and compact substructures, leaving out those relatively scattered GSE stars.

To sum up, we believe that energy is a highly significant quantity for understanding the internal structure of GSE, so we divide GSE into three parts according to energy.
Firstly, a prominent high-energy GSE group (hot pink) is separately regarded as one part because it is also isolated in the $(r_{\rm apo},\ e)$ space, as shown in the top-left panel of Figure \ref{figgroups_errorbar}.
Around $r_{\rm rapo}\sim10\,{\rm kpc}$, the GSE groups assemble into a large clump in the $(r_{\rm apo},\ e)$ space, corresponding to a major clump with $E\,\textless\,-1.5\,[\times10^5\ {\rm km^2\,s^{-2}}]$ in the $(E,\ L_z)$ space that would transform into a smooth asymmetric peak found by \citet{Belokurov2023MNRAS} if the dispersed and smooth GSE halo stars in this space are considered.
In order to ensure sufficient stars in the medium and low energy parts and to understand the distinct properties on both sides of the peak, we divide the clump or peak around $r_{\rm apo}\sim10\,{\rm kpc}$ or $E\sim-1.7\,[\times10^5\ {\rm km^2\,s^{-2}}]$ into two parts (pink and cyan circles represent the medium and low energies, respectively).
A slightly retrograde group (royal blue) has a mean eccentricity greater than 0.7 and lies on the boundary between GSE and Thamnos in the 4D space, so we consider it as an individual GSE part for comparison with the major GSE and Thamnos.

In Figure \ref{figgroups_errorbar}, the eccentric GSE groups are located at the bottom of $(J_z-J_R)/J_{\rm tot}$ and $J_{\phi}/J_{\rm tot}$ space, indicating that GSE stars move on extremely radial orbits. Their pericentric distances aggregate around $r_{\rm peri}\approx1$\,kpc, but the GSE stars definitely do not build up here because their radial velocities reach their maximal values when the stars on eccentric orbits approach their pericentres.
\citet{naidu2021} and \citet{Belokurov2023MNRAS} have both successfully reproduced the generic features of GSE in phase space and spatial position through simulations of radial accretions of massive galaxies, despite discrepancies concerning stripped GSE's outer disc.
While the overall characteristics of GSE are widely recognized, it is also worthwhile to study its internal structure, as we analyze below.
In Figure \ref{figgroups_errorbar}, we found that the high-energy GSE (hot pink) stars constitute an obvious apocenter pile-up in the range of $r_{\rm apo}=29.5\pm3.6\,{\rm kpc}$ and move on orbits with strongly radial action ($(J_{z}-J_{R})/J_{\rm tot}=-0.89\pm0.11$, $J_R=2199.86\pm324.72\,{\rm kpc\,km\,s^{-1}}$) and high eccentricity ($e=0.93\pm0.04$).
Therefore, the high-energy GSE stars could be responsible for the break in a broken power-law density profile at $r\approx28-30\,{\rm kpc}$ \citep{naidu2021,Han2022,Ye2023MNRAS} that also is predicted by chevrons in the $r\text{--}v_r$ space \citep{Belokurov2023MNRAS,Wu2023ApJ}.
Instead, the apocentric distance of the low-energy GSE (cyan) is the smallest, and its orbit has relatively low eccentricity ($e=0.84\pm0.07$) and radial action ($(J_{z}-J_{R})/J_{\rm tot}=-0.71\pm0.17$, $J_R=569.81\pm83.42\,{\rm kpc\,km\,s^{-1}}$) compared to the high/medium-energy GSE.
The medium-energy GSE (pink) part with the largest proportion has an intermediate apocentric distance, eccentricity and radial action ($r_{\rm apo}=13.0\pm2.7\,{\rm kpc}$, $e=0.88\pm0.06$, $(J_{z}-J_{R})/J_{\rm tot}=-0.84\pm0.12$, $J_R=921.27\pm272.08\,{\rm kpc\,km\,s^{-1}}$), so they could be associated with the apocenter pile-up that creates the break at 15-18\,kpc \citep{naidu2021}.
For the three GSE parts with different energy levels, we suggest that the part with higher energy has a farther apocentric distance and larger eccentricity but a smaller $(J_{z}-J_{R})/J_{\rm tot}$ and radial action $J_R$, which indicates that its members move on more radial orbits, because chevrons in the $r\text{--}v_r$ space show high radial velocities for the high-energy GSE stars. 

In Figure \ref{figallgroups}, we give all the mergers in dynamical space for easy comparison of their dynamical properties.
The maximum vertical height of high-energy GSE part (hot pink circle) has very large uncertainty, because the high-energy GSE group (hot pink circle) is relatively scattered in the action-energy space, as shown in Figure \ref{figallgroups}, especially $J_z$ and $J_R$, but it is significantly higher than those of medium and low due to much farther apocentre.
The slightly retrograde GSE part has a somewhat larger pericentric distance than those of the other parts but smaller than that of Thamnos, so it could belong to the GSE merger.
Figure \ref{figallgroups_v} shows the 3D kinematic properties of GSE (hot pink, pink, cyan, and royal blue circles). 
The major difference of distributions of velocity for three distinct energy GSE (hot pink, pink, cyan) is about radial velocity $v_R$. 
Note that the cylindrical coordinate system is used here to compare with action \textbf{\textit{J}}.
The slightly retrograde GSE part (royal blue) is positioned between the other three GSE parts and Thamnos in the action-energy space or $(v_R,\ v_{\phi})$ space, so it could be a mixture of the GSE and Thamnos.
However, the GSE members could be dominated due to their high eccentricities ($e=0.73\pm0.06$).

In order to understand the correlation between the eccentricity and velocity for the four GSE parts, we also show the GSE member stars in the velocity-eccentricity space in Figures \ref{figallgroups_evr} and \ref{figallgroups_evt}.
Note that the spherical coordinate system is used here to understand the dependence of radial ($v_r$) and tangential ($\sqrt{v_{\phi}^2+v_{\theta}^2}$) velocities on eccentricity. 
We found that the eccentricities of GSE stars with high radial velocities slowly decreases with $\lvert v_{\phi}\rvert$,
because, for a star moving with high radial velocity $v_r$ at low zenithal angles, its orbit tends to be radial, while fast rotation $(v_{\phi},\ v_{\theta})$ makes its shape more circular under the assumption that they orbit the Galactic centre.
Since there is no significant difference in the zenithal velocity for the four parts, we can only determine the influence of $v_r$ and $v_{\phi}$ on eccentricity for a given level of $v_{\theta}$ in the solar neighborhood.
We conclude that, for the local GSE, the energy level is positively correlated with eccentricity or radial degree of orbit, and vice versa for $v_{\phi}$.

\subsection{Thamnos}\label{Thamnos}

\citet{koppelman2019multipleretrograde} suggested the existence of a significant, retrograde substructure, which could be divided into two parts on the basis of metallicity and azimuthal velocity, named Thamnos\,1 and Thamnos\,2.
We select four groups associated with Thamnos from the residual 10 groups.
In order to study the fine dynamical structures of Thamnos, we split it into two parts: $(J_{z}-J_R)\,\textless\,0$ (orange square) and $(J_{z}-J_R)\,\textgreater\,0$ (blue square), as shown in Figures \ref{figgroups_errorbar}-\ref{figallgroups_evt}.
When comparing the two parts of Thamnos, we found no peculiar difference in energy, pericentric distance, apocentric distance and eccentricity, while significant differences in azimuthal velocity ($v_{\phi}$) and vertical velocity ($v_z$), leading to their separation in the $J_z\text{--}J_{\phi}$ space as shown in Figure \ref{figallgroups}.
The distribution of the members of the Thamnos part (blue square) with $(J_z-J_R)/J_{\rm tot}\textgreater0$ in $(J_z-J_R)/J_{\rm tot}$ vs. $J_{\phi}/J_{\rm tot}$ space is consistent with that of \citet{naidu2020}.

\citet{koppelman2019multipleretrograde} found that 
Thamnos\,2 embraces Thamnos\,1 in the $v_R\text{--}v_z$ space but Thamnos\,1 has higher energy and azimuthal velocity than Thamnos\,2.
In the top-right panel of Figure \ref{figallgroups_v}, we found that the Thamnos part with $(J_z-J_{\phi})/J_{\rm tot}\,\textgreater\,0$ cannot fully embrace another Thamnos part with $(J_z-J_{\phi})/J_{\rm tot}\,\textless\,0$ in the $v_R-v_{z}$ space, and that their azimuthal velocities $(v_{\phi}=84.84\pm24.88\,{\rm km\,s^{-1}},\,120.07\pm34.14\,{\rm km\,s^{-1}})$ are lower than those $(v_{\phi}\sim150\,{\rm km\,s^{-1}},\,200\,{\rm km\,s^{-1}})$ of \citet{koppelman2019multipleretrograde}, which might be caused by their clustering in a different 4D space (i.e., $E$, $L_z$, $e$ and [Fe/H]) to probe the high-density regions of Thamnos and other differences.
\citet{koppelman2019multipleretrograde} divide Thamnos into two parts with the different metallicities and azimuthal velocities in the $E\text{--}L_z$ space, and do not clearly classify stars at low energies ($E\,\textless\,-1.5\,[\times10^5\ {\rm km^2\,s^{-2}}]$) and $\lvert L_z\rvert$, and we split it into more polar and radial parts in the projected action space, with distinct vertical and azimuthal velocities.
In addition, they centred the clustering results from HDBSCAN \citep{hdbscan2017} and set boundary lines in the $E\text{--}L_z$ space to select Thamnos stars, while we directly used the clustering results from SNN as Thamnos stars because we are not sure whether the smooth stellar halo stars in these areas belong to Thamnos, maybe they are the slightly retrograde GSE stars.
Finally, the Thamnos stars in the range of $L_z\sim[1200,\,1600]\,{\rm kpc\,km^{-1}}$ is actually Thamnos 1 proposed by \citet{koppelman2019multipleretrograde}.

Below we analyze the differences between polar and radial Thamnos parts.
We know that the maximum vertical height ($z_{\rm max}$) is related to vertical velocity ($v_{z}$), that is, a star with large $v_z$ usually can reach a high $z_{\rm max}$, so the polar Thamnos (blue square) stars move along orbits reaching higher $z_{\rm max}$.
In the $e\text{--}v_{\theta}$ space, for polar Thamnos, it is evident that $e$ decreases as $\lvert v_{\theta}\rvert$, whereas for radial Thamnos, there appears to be no significant correlation with zenithal velocity, somewhat resembling the behaviour observed in the GSE.
However, in the $e\text{--}v_{\phi}$, they appear similar descending characteristics, with the $e\text{--}v_{\phi}$ slope of radial Thamnos (orange square) being comparable to that of the retrograde GSE (royal blue circle), while that of polar Thamnos (blue square) is significantly steeper.
Within $e\text{--}v_r$ space, radial and polar Thamnos parts have similar distributions, their radial velocities differ significantly from GSE because they are better suited to a single Gaussian rather than a mixture of two Gaussians describing radial anisotropic GSE \citep{Lancaster2019MNRAS,Necib2019ApJ,Iorio2021MNRAS,wuwenbo2022} for radial velocity ($v_r$).
Compared with the GSE, Thamnos is a retrograde component with low radial velocity and action, so its member stars move along orbits with smaller eccentricity $e=0.51\pm0.15$.
Furthermore, we notice that the slightly retrograde GSE part is very close to Thamnos in phase spaces, recent studies \citep[e.g.,][]{koppelman2020A&Anbody,naidu2021} show that the GSE stars lost early have large retrograde motions and a subset of these stars have low eccentricities, so these retrograde populations, such as Thamnos, Arjuna and Sequoia, may come from the same progenitor galaxy as GSE itself, which will require observational details and high-resolution simulations to be further confirmed in the future.

\subsection{Metal-weak Thick Disk and Helmi Streams}

Here, we select the Metal-weak Thick Disk \citep[MWTD,][]{carollo2019ApJ} group based on its dynamical properties described by \citet{naidu2020} in $(J_z-J_R)/J_{\rm tot}$ vs. $J_{\phi}/J_{\rm tot}$ space.
However, we found that the VMP stars of the MWTD have larger eccentricities ($e=0.61\pm0.07\,\textgreater\,0.47$) than those of more metal-rich ($\rm [Fe/H]\sim-1.12$) MWTD members in \citet{naidu2020} even if we add the criterion of $\rm[\alpha/Fe]$ in \citet{naidu2020}, namely $0.25\,\textless\,[\rm\alpha/Fe]\,\textless\,0.45$, which suggests that the orbital properties of the MWTD may be strongly correlated with chemical abundance, and the VMP part of the MWTD may originate from accreted galaxies, which require high-completeness samples for further verification.
After removing the MWTD ($v_{\phi}=-142.30\pm28.51\,{\rm km\,s^{-1}}$) from the two extremely prograde groups, the only left group ($v_{\phi}=-143.95\pm27.86\,{\rm km\,s^{-1}}$) could belong to the Helmi Streams, because its actions and energy, as listed in Table \ref{taballmerger}, are similar to the Helmi Streams obtained by \citet{yuandynamicalrelics2020} using VMP stars.

The discovery of the Helmi Streams as accreted substructures in the halo was one of the first instances achieved via integrals of motion due to their high vertical velocities \citep{helmi1999}.
In order to further make sure that the group belongs to the Helmi Streams, we use the orthogonal and $z$ components of the angular momentum (i.e., $L_{\bot}=\sqrt{L_{x}^{2}+L_{y}^{2}}$, $L_z$) to identify the Helmi Streams, similar to \citet{koppelman2019multipleretrograde} and \citet{naidu2020}.
The Helmi Streams selected here are satisfied with the selection criteria of \citet{naidu2020}, namely $-1.7\,\textless\,L_z/[{\rm 10^3\,kpc\,km\,s^{-1}}]\,\textless\,-0.75$ and $1.6\,\textless\,L_{\bot}/[{\rm 10^3\,kpc\,km\,s^{-1}}]\,\textless\,3.2$, as shown in the bottom-right panel of Figure \ref{figallgroups_v}, and are of higher energy than the MWTD.
Although both the Helmi Streams and MWTD are highly prograde and are adjacent in the $(E,\,J_{\phi})$ and $(J_R,\,J_{\phi})$ spaces, the vertical velocities and actions of the Helmi Streams are much larger than those of MWTD.
Therefore, the MWTD stars move along radial and relatively eccentric orbits near the Galactic disc, and the Helmi Streams appear to extend out of the Galactic disc, which indicates that the progenitor galaxy of Helmi streams, with a low initial tangential velocity, might accrete with the Milky Way in a direction nearly perpendicular to the Galactic disc.
In addition, the MWTD has a richer $\rm [\alpha/Fe]$ than that of Helmi Streams ($\rm [\alpha/Fe]_{MWTD}=0.29\pm0.17$, $\rm [\alpha/Fe]_{Helmi\ Streams}=0.21\pm0.16$), indicating they do not appear to originate from a common progenitor galaxy.

\subsection{Wukong}

It is found by \citet{naidu2020} and \citet{yuan2020lowmass} that Wukong/LMS-1 is a less prograde and more metal-poor merger compared to the Helmi Streams, and its member stars also move along polar orbits due to their large vertical actions.
\citet{malhan2022} apply globular clusters, dwarf galaxies, and streams as tracers to probe accretion events and also determine that Wukong has a slight prograde motion and its objects have very polar orbits, and that there are three most metal-poor streams of our galaxy belonging to Wukong.
Therefore, we regard the only two left groups with slightly prograde rotation as Wukong, and the distributions of their member stars in $(J_z-J_R)/J_{\rm tot}$ vs. $J_{\phi}/J_{\rm tot}$ space and $(E,\ J_{\phi})$ spaces are very consistent with those in \citet{naidu2020}.
Similar to the GSE, we divide Wukong into two parts with an energy difference of $\sim0.2\times10^{5}\,{\rm km^2\ s^{-2}}$ to understand its internal structure under distinct energies.
Two Wukong parts are not satisfied with the selection criteria of the Helmi Streams \citep{naidu2020}, due to their low $z$ component of angular momentums, and Wukong can reach a vertical height much higher than that for MWTD. Therefore, we can exclude the possibility that they belong to the Helmi Streams and MWTD.

Below we analyze the respective features and relationship between high (purple up triangle) and low (dark cyan up triangle) energy Wukong parts. Their $J_{\phi}\text{--}J_R$ distributions almost overlap, but a significant discrepancy is in vertical action due to distinct absolute vertical or polar velocity ($\lvert v_z\rvert$ or $\lvert v_{\theta}\rvert$).
Wukong overlaps with the prograde GSE in $E\text{--}L_z$ space, but their prominent $z_{\rm max}$ and $J_z$ as well as unremarkable $e$ all imply that, observationally, it is very different from GSE.
However, the massive GSE progenitor merged with the Milky Way motivates the significant motion of the host galaxy, which could disrupt the precession of the angular momentum of the host disc, and after the merger, it may continue to precess. Therefore, we cannot entirely rule out the probability that the Wukong progenitor comes from a metal-poor area of GSE.

Finally, we combine the aforementioned GSE and Thamnos to discuss the correlation between pericentric distance and rotation velocity.
We have found that the slightly retrograde GSE part (royal blue) possesses a farther pericentric distance compared to other GSE parts of distinct energy levels.
Here, the Wukong part (purple up triangle) with a larger $\lvert v_{\theta}\rvert$ also shows a larger pericentric distance (i.e., $r_{\rm peri, purple}=6.05\pm1.44\,{\rm kpc},\ r_{\rm peri, darkcyan}=2.83\pm0.79\,{\rm kpc}$). 
The stars of the Thamnos part (orange) with smaller $v_{\theta}$ but larger $v_{\phi}$ possess pericentric distances as large as another part (blue square), so pericentric distance usually increases with rotation velocity.
\subsection{Highly retrograde I'itoi+Sequoia+Arjuna and Pontus}

\begin{table*}
  \caption{Summary of existing substructures in local VMP halo.}
  \label{taballmerger}
  \adjustbox{scale=0.9}{
  \begin{tabular}{rrrrrrrrrrr}
    \hline
    Substructure & $(N_{\rm sub})^{a}$ & $\rm [\alpha/Fe]$ & $e$ & $r_{\rm apo}$ & $r_{\rm peri}$ & $z_{\rm max}$ & $(J_z-J_R)/J_{\rm tot}$ & $J_{\phi}/J_{\rm tot}$ & $E$ & $J_R$\\
     &  &  &  & (kpc) & (kpc) & (kpc) &  &  & $({\rm10^5\, km^2\,s^{-2}})$ & (${\rm kpc\,km\,s^{-1}}$) \\
    \hline
    Gaia-Sausage/Enceladus & 1070 & 0.263 & 0.86 & 12.36 & 0.87 & 4.45 & -0.78 & -0.02 & -1.63 & 840.2\\
    Thamnos & 274 & 0.261 & 0.51 & 9.87 & 3.31 & 3.70 & -0.08 & 0.88 & -1.65 & 273.9\\
    Metal-weak Thick Disk & 80 & 0.287 & 0.61 & 14.58 & 3.46 & 4.99 & -0.28 & -0.89 & -1.48 & 552.4\\
    Helmi Streams & 45 & 0.205 & 0.37 & 17.74 & 8.18 & 15.05 & 0.55 & -0.68 & -1.32 & 252.5\\
    Wukong & 120 & 0.239 & 0.48 & 10.92 & 4.04 & 10.04 & 0.60 & -0.33 & -1.58 & 236.2\\
    I'itoi/Sequoia/Arjuna & 47 & 0.249 & 0.59 & 23.85 & 5.96 & 10.45 & -0.23 & 0.87 & -1.25 & 840.6\\
    Pontus & 67 & 0.287 & 0.74 & 8.81 & 1.39 & 7.22 & 0.03 & 0.26 & -1.73 & 385.0\\
    \hline
    \hline
    $J_z$ & $J_{\phi}$ & $v_r$ & $v_{\theta}$ & $v_{\phi}$ & $\sigma_{v_r}$ & $\sigma_{v_\theta}$ & $\sigma_{v_\phi}$ & $\lvert v_{r}\rvert$ &$\lvert v_{\theta}\rvert$ & $\lvert v_{\phi}\rvert$ \\
    (${\rm kpc\,km\,s^{-1}}$) & (${\rm kpc\,km\,s^{-1}}$) & ($\rm km\,s^{-1}$) & ($\rm km\,s^{-1}$) & ($\rm km\,s^{-1}$) & ($\rm km\,s^{-1}$) & ($\rm km\,s^{-1}$) & ($\rm km\,s^{-1}$) & ($\rm km\,s^{-1}$) & ($\rm km\,s^{-1}$) & ($\rm km\,s^{-1}$)\\
    \hline
    87.7 & -23.0 & -14.2 & 2.0 & -3.3 & 168.1 & 50.4 & 42.0 & 147.9 & 41.2 & 35.3\\
    196.1 & 937.4 & -7.6 & 1.2 & 111.6 & 87.0 & 83.8 & 35.5 & 73.8 & 66.8 & 111.6\\
    171.2 & -1181.2 & -35.2 & 15.2 & -142.3 & 184.9 & 81.8 & 28.5 & 185.9 & 72.1 & 142.3\\
    1276.7 & -1251.4 & 1.5 & 216.2 & -144.0 & 78.5 & 133.5 & 27.9 & 57.1 & 253.0 & 144.0\\
    1014.2 & -308.8 & 0.09 & -38.6 & -39.2 & 92.6 & 181.7 & 31.1 & 83.5 & 178.9 & 42.5\\
    342.2 & 1906.4 & -8.0 & -12.2 & 229.7 & 186.5 & 135.7 & 55.7 & 169.4 & 120.9 & 229.7\\
    422.8 & 177.4 & 2.3 & 13.3 & 23.5 & 69.4 & 112.4 & 28.7 & 54.7 & 111.0 & 27.4\\
    \hline
    \multicolumn{3}{l}{$^a$ $N_{\rm sub}$ represents the number of stars in substructure.}\\
  \end{tabular}
  }
\end{table*}
I'itoi \citep{naidu2020}, Sequoia and Arjuna \citep{myeong2019} (ISA) are three highly retrograde and high-energy populations with distinct chemical abundances (i.e., $\rm [\alpha/Fe],\ [Fe/H]$).
Therefore, we take the only extremely retrograde and high-energy group as ISA.
Although ISA and Thamnos overlap in $(J_z-J_R)/J_{\rm tot}$ and $J_{\phi}/J_{\rm tot}$ space, the energy, retrograde rotation and apocentric distance are much larger than those of Thamnos.
Its remarkable energy is consistent with that of Sequoia in Figure 2 of \citet{koppelman2019multipleretrograde} as shown in the bottom-right panel of Figure \ref{figallgroups} here.
Based on three metallicity levels, \citet{naidu2020} clearly defined the Arjuna stars as those of $\rm[Fe/H]\,\textgreater\,-1.5$, the Sequoia stars of $\rm -2\,\textless\,[Fe/H]\,\textless\,1.5$, and the I'itoi stars of $\rm [Fe/H]\,\textless\,-2$.
The bottom-left panel of Figure \ref{figallgroups} shows that a few ISA members satisfy $(J_z-J_R)\,\textgreater\,0$, while \citet{naidu2020} found that Arjuna and Sequoia obviously satisfy this condition (for Arjuna and Sequoia $(J_z-J_R)/J_{\rm tot}=0.16$, for I'itoi $(J_z-J_R)/J_{\rm tot}=0.09$).
Considering that the tracers in this work are VMP stars with $\rm [Fe/H]\,\textless\,-1.8$ and the member stars of Arjuna and Sequoia are mostly located outside the solar neighbourhood ($r\sim22.91\,{\rm kpc},\,\lvert z\rvert\sim16.66\,{\rm kpc}$ for Arjuna, $r\sim15.55\,{\rm kpc},\,\lvert z\rvert\sim11.02\,{\rm kpc}$ for Sequoia, and $r\sim12.37\,{\rm kpc},\,\lvert z\rvert\sim7.46\,{\rm kpc}$ for I'itoi, see Table 1 of \citet{naidu2020} for details), the ISA group is composed of a lot of ultra-metal-poor stars from I'itoi and a small number of Sequoia stars as well as very few Arjuna stars.
Here the ISA stars could be more retrograde than Sequoia in other works \citep[e.g.,][]{Feuillet2021MNRAS} because we pursue purity and discover its member stars through clustering, while previous studies wish to get high-completeness member stars by actions \textit{\textbf{J}} cuts, which actually overlap with our ISA members in projected action space. In addition, the retrograde halo is more metal-poor, as advocated by \citet{koppelman2019multipleretrograde}.

Pontus recently discovered by \citet{malhan2022} is a slightly retrograde merger, and it is indistinguishable from the GSE in ($E,\ L_z$) space or ($v_r,v_{\phi}$) space.
Since stars with the same origin have a common age-metallicity relationship \citep{massari2019origin}, \citet{malhan2022} ruled it out as a fragment of the GSE by comparing its age-metallicity relationship with that of the GSE.
Its dynamical properties are $E=-172842\pm3477\,{\rm km^2\,s^{-1}}$, $J_R=385\pm89\,{\rm km\,s^{-1}\,kpc}$, $J_{\phi}=177\pm214\,{\rm km\,s^{-1}\,kpc}$, $J_z=423\pm128\,{\rm km\,s^{-1}\,kpc}$, $L_{\bot}=896\pm154\,{\rm km\,s^{-1}\,kpc}$, $e=0.74\pm0.10$, $r_{\rm apo}=8.8\pm0.7\,{\rm kpc}$, $r_{\rm peri}=1.4\pm0.6\,{\rm kpc}$ and $z_{\rm max}=7.2\pm1.0\,{\rm kpc}$.
Figure \ref{figallgroups} shows that Pontus is located at a distinct position from the GSE and Thamnos in the projected action space, which indicates that the members of the GSE move along more in-plane and radial orbits, but Thamnos is more retrograde compared to Pontus.
\section{Discussion and Conclusions}\label{Conclusion}

Based on the precise parallaxes and proper motions from \textit{Gaia} DR3 in combination with the radial velocities, $\rm[\alpha/Fe]$ and [Fe/H] from LAMOST DR9,
we select the VMP stars (i.e., $\rm [Fe/H]\,\textless\,-1.8$) to explore the dynamical relics of halo objects in the solar neighbourhood. 
Since small or low-mass systems are primarily populated by very metal-poor stars, the VMP sample can help us to trace fine structures that would be overwhelmed during clustering if all samples were considered.
In this work, we apply the improved two-stage SNN clustering algorithm to identify substructures with similar actions and energy ($\textbf{\textit{J}},\ E$).
Not only the observation errors are considered, but also the 10,\,000 SNN clustering results are further clustered to obtain stable groups in order to overcome the parameter adjustment problem.
In total, we identify 26 groups and associate them with seven known mergers based on their kinematic and dynamical properties, including Gaia-Sausage-Enceladus, Thamnos, Wukong, Metal-weak Thick Disk, Helmi Streams, I'itoi+Sequoia+Arjuna, and Pontus.
Their fractions are as follows: Gaia-Sausage-Enceladus ($62.83\%$), Thamnos ($16.09\%$), Wukong ($7.05\%$), Metal-weak Thick Disk ($4.70\%$), Pontus ($3.93\%$), I'itoi+Sequoia+Arjuna ($2.76\%$) and Helmi Streams ($2.64\%$).
We did not find unknown substructures, perhaps, because our sample is limited to the solar neighborhood.

In order to obtain the characteristics of each merger in more detail, we further investigate the parts of distinct dynamical properties (i.e., $E$ and $(J_z-J_R)/J_{\rm tot}$) for the GSE, Thamnos, and Wukong.
In energy and angular momentum $(E,\,L_z)$ space, the GSE stars appear slightly prograde at high energies but symmetric with respect to $\lvert L_z \rvert=0$ at low energies. The $E$ distribution is also rather wrinkly with several overdensities gathered in low $\lvert L_z \rvert$. Similarly, it corresponds to over-dense chevrons in the $r\text{--}v_r$ phase-space.
The high-energy and medium-energy GSE parts, with apocenters in the ranges of $29.5\pm3.6\,{\rm kpc}$ and $13.0\pm2.7\,{\rm kpc}$, presumably contribute to the two breaks at $15-18\,{\rm kpc}$ and $30\,{\rm kpc}$ in the ``double-break'' density profile predicted by \citet{naidu2021}, respectively.
We found that the difference between two slightly retrograde Thamnos parts, with energy in the same range of $-165280\pm4171{\rm\,km^{2}\,s^{-2}}$, is primarily due to their zenithal or vertical and azimuthal velocities, which results in two clumps (i.e., $(J_{z}-J_R)/J_{\rm tot}\,\textless\,0$ and $(J_{z}-J_R)/J_{\rm tot}\,\textgreater\,0$) in the projected action space.
The difference between two Wukong parts in vertical action or velocity leads to their slightly distinct orbital parameters such as $z_{\rm max}$ and $r_{\rm apo}$.
The VMP members of MWTD move along the orbits with larger eccentricities than those of more metal-rich ($\rm [Fe/H]\sim-1.12$) the MWTD in \citet[][]{naidu2020}, so we suggest that the VMP part of the MWTD may originate from accreted galaxies.
Orbits of the highly prograde Helmi Streams in local samples are circular and rise to high vertical heights.
We also analyze the dynamical properties of highly retrograde ISA with high energy, and Pontus hidden in the GSE, but we found that the members of Pontus move on more polar orbits than those of the GSE stars.
In Table \ref{taballmerger} we summarize all the mergers explored in this study that constitute the VMP halo in the solar neighbourhood.

Our work reveals the origin of local metal-poor stellar halos, which may provide significant insights into probing the origins of exotic stars, such as high-velocity stars.
Study in the origin of high-velocity stars, such as the tidal debris of disrupted dwarf galaxies \citep{Du2018ApJhighvelocity,Du2018ApJLorigin} and LMC \citep{Erkal2019MNRAS}, is very interesting, recent results \citep[e.g.,][]{huang2021ApJL,li2022ApJL,Liqingzheng2023AJ} show some candidate high-velocity stars originating from the Sgr dwarf spheroid galaxy.
It is known that the energies of the Sgr member stars are so high that some member stars satisfy $E\,\textgreater\,0$ or $v\,\textgreater\,v_{\rm escape}$ (escape velocity).
In the 26 groups we identified here, two special groups, namely high-energy GSE (hot pink circle) and ISA (red diamond), were discussed further due to their high velocities and energies, with their velocities mostly exceeding $300\,{\rm km\,s^{-1}}$.
As we expected, stars at very high energies are so sparse that they can hardly be identified as clumps in the action-energy space and the two high-energy groups are only about $E=-1.2\,[\times10^5\ {\rm km^2\,s^{-2}}]$.
The retrograde ISA stars may come from the GSE's outer disc due to the host galaxy moving significantly in the merger.
Therefore, we conjecture that some VMP high-velocity stars, especially those with large radial or extremely retrograde velocities, may originate from the massive GSE satellite, and its orbital angular momentum tilted by $30^{\circ}$ from that of the host disc when merging \citep{naidu2021}.

\section*{Acknowledgements}
We thank the referee for the insightful comments and suggestions, which have improved the paper significantly.
This work was supported by the National Natural Sciences Foundation of China (NSFC Nos: 12090040, 12090044, 11973042, 11973052 and 11873053.)
It was also supported by the Fundamental Research Funds for the Central Universities and the National Key R\&D Program of China No. 2019YFA0405501. 
This work has made use of data from the European Space Agency (ESA) mission
{\it Gaia} (\url{https://www.cosmos.esa.int/gaia}), processed by the {\it Gaia}
Data Processing and Analysis Consortium (DPAC,
\url{https://www.cosmos.esa.int/web/gaia/dpac/consortium}). Funding for the DPAC
has been provided by national institutions, in particular, the institutions
participating in the {\it Gaia} Multilateral Agreement.
Guoshoujing Telescope (the Large Sky Area Multi-Object Fiber Spectroscopic Telescope LAMOST) is a National Major Scientific Project built by the Chinese Academy of Sciences. Funding for the project has been provided by the National Development and Reform Commission. LAMOST is operated and managed by the National Astronomical Observatories, Chinese Academy of Sciences.

\section*{Data Availability}
This study uses publicly available \textit{Gaia} DR3 and LAMOST DR9 data.



\bibliographystyle{mnras}
\bibliography{yds.bbl} 





\bsp	
\label{lastpage}
\end{document}